\documentstyle[12pt,epsf,floats,prl,aps]{revtex}    
\newcommand{\figwidth}{3.375in}     
\begin{document}
\draft
\title{Critical behavior of the planar magnet model in three dimensions}   
\author{Kwangsik Nho and Efstratios Manousakis}
\address{Department of Physics and Center for Research and Technology,\\
Florida State University, Tallahassee, Florida 32306-4350}
\date{\today}
\maketitle
\begin{abstract}
We study  the critical behavior of the three-dimensional planar magnet 
model in which each spin is considered to have three components of 
which only the $x$ and $y$ components  are coupled. 
We use a hybrid Monte Carlo algorithm in which a single-cluster 
update  is combined with the over-relaxation  and 
Metropolis spin re-orientation algorithm. Periodic boundary conditions were 
applied in all directions. We have calculated the fourth-order cumulant  
in finite size lattices using the single-histogram 
re-weighting method. Using  finite-size scaling theory, 
we obtained the critical temperature which is very different from that of the
usual $XY$ model. At the critical temperature, we calculated 
the susceptibility and the magnetization on lattices of size up to $42^3$.
Using finite-size scaling theory we accurately determine the critical exponents of the model and find that 
$\nu$=0.670(7), $\gamma/\nu$=1.9696(37), and $\beta/\nu$=0.515(2).
Thus,  we conclude that the model belongs to the same universality class 
with the $XY$ model, as expected.
\end{abstract}


\section{Introduction}
\label{sec0}
Our understanding of critical phenomena has been significantly advanced
with the development of the renormalization-group (RG) theory\cite{Wilson}. 
The RG theory predicts relationships between groups of exponents and that there is 
a universal behavior. In a second order phase transition, the correlation length $\xi$ 
diverges as  the critical point is approached, and so
the details of the microscopic Hamiltonian are unimportant for the 
critical behavior.
All members of a given universality class have identical critical behavior and 
critical exponents.

The three-dimensional classical XY model is relevant to the critical 
behavior of many physical systems, such as superfluid 
$^{4}He$,  magnetic materials and the high-Tc superconductors.
In the pseudospin notation, this model is defined by the Hamiltonian
\begin{equation}
H=-J\sum_{<ij>}(S_{i}^{x}S_{j}^{x}+S_{i}^{y}S_{j}^{y}),
\label{XY}
\end{equation}
where the summation is over all nearest neighbor 
pairs of sites $i$ and $j$ on a simple cubic lattice. In this model one considers that the
spin has two components, $\vec S_{i}= (S_{i}^x,S_{i}^y)$ and $S_i^{x 2}+S_i^{y 2}=1$. 

In this paper we wish to consider
a three component local spin $\vec S_{i}= (S^x_i,S^y_i,S^z_i)$ and the same Hamiltonian as given by
Eq. (\ref{XY}) (namely, with no coupling between the z-components of the spins) in 
three dimensions.  Even though the Hamiltonian is the same,
namely, there is no coupling between the z-component of the spins, the constrain for 
each spin is $(S_i^{x})^2 + (S_i^{y})^2+(S_i^{z})^2=1$, which implies that 
the quantity $(S_i^{x})^2+(S_i^{y})^2$ is
fluctuating. In order to be distinguished from the usual XY model,
the name {\it planar magnet model} will be adopted for this model.

The reason for our desire
to study this model is that it is related directly to the so-called model-F\cite{Hohenberg} 
used to study non-equilibrium phenomena in systems, such as superfluids, 
with a two-component order parameter and a conserved current.
In the planar magnet model, the order parameter is not a constant of the motion.
A constant of the motion is the $z$ component of the 
magnetization. Thus, there is  an important 
relationship between the order parameter and the $z$ component of magnetization,
which is expressed by a Poisson-bracket relation\cite{Hohenberg}. 
This equation is crucial for the hydrodynamics and the critical dynamics of the system.
One therefore needs to find out the critical properties of this model in order
to study non-equilibrium properties of superfluids or other systems described 
by the model F. In future work, we shall use model-F 
to describe the dynamical critical phenomena of superfluid helium.
Before such a project is undertaken, the static critical properties of the
planar magnet model should be investigated accurately. 

Although the static properties of the $XY$ model with $\vec{S}_{i}=
(S^{x}_{i},S^{y}_{i})$ have been investigated by a variety of
statistical-mechanical 
methods\cite{Guillou,Albert,high,Li,MC,S2,S3,Janke,Hasenbusch}, 
the system with $\vec{S}_{i}=
(S^{x}_{i},S^{y}_{i},S^{z}_{i})$  has been given much less attention.
So far the critical behavior of this model has been studied
by high temperature expansion\cite{Ferer} and Monte Carlo(MC) simulation methods\cite{Costa,Oh}.
High temperature expansion provides the value for the critical temperature and the critical 
exponents. In these recent MC calculations\cite{Costa,Oh}, only the critical 
temperature is determined.  These MC calculations were carried out on  small size systems and thus only rough estimates are available.  

In this paper we study the three-dimensional planar magnet model
 using a hybrid Monte Carlo method
(a combination of the cluster algorithm
with over-relaxation and Metropolis spin re-orientation algorithm) in conjuction 
with single-histogram re-weighting technique and finite-size scaling. 
We calculate  the  fourth order cumulant,
the magnetization, and the susceptibility (on cubic lattices $L\times L \times L$ with 
$L$ up to $42$) and from 
their finite-size scaling behavior 
we determine the critical properties of the 
planar magnet model accurately.

\section{Physical Quantities and Monte Carlo Method}
\label{sec1}
Let us first summarize the
definitions of the observables that are calculated in our simulation.
The energy density of our model is given by
\begin{equation}
<e>=E/V=\frac{1}{V}\sum_{<ij>}<S_{i}^{x}S_{j}^{x}+S_{i}^{y}S_{j}^{y}>,
\end{equation}
where $V=L^{3}$ and the angular brackets denote the thermal average.
The fourth-order cumulant $U_{L}(K)$\cite{Binder} can be written as
\begin{equation}
U_{L}(K)=1-\frac{<m^{4}>}{3<m^{2}>^{2}},
\end{equation}
where $m=\frac{1}{V}(M_{x}^{2}+M_{y}^{2}+M_{z}^{2})^{1/2}$ is the magnetization per spin,
$\vec M = \sum_{i} \vec S_i$
 and $K=J/(k_{B}T)$ is the coupling, or the reduced inverse temperature
in units of $J$. The fourth-order cumulant $U_{L}(K)$ is one important quantity which we use to
determine the critical coupling constant $K_{c}$.
 In the scaling region close to the critical coupling, the fourth-order 
cumulant $U_{L}(K)$ as function of $K$ for different values of $L$ are lines which 
go through the same point. 

The magnetic susceptibility per spin $\chi$ is given by
\begin{equation}
\chi = VK(<m^{2}>-<\vec{m}>^{2}),
\end{equation}
where $\vec{m}$ is the magnetization vector per spin.

The three-dimensional planar magnet model with ferromagnetic interactions $J>0$ has a second-order phase transition.
In simulations of  systems near a second-order phase transition, a major difficulty
arises which is known as critical slowing down. The critical slowing down can be reduced 
by using several techniques and what we found as optimal for our case 
was to use the hybrid Monte Carlo algorithm as described in Ref. \cite{Landau}.
Equilibrium configurations were created using a hybrid Monte Carlo algorithm 
which combines cluster updates of in-plane spin components\cite{Wolff} with Metropolis 
and over-relaxation\cite{Brown} of spin re-orientations. After each single-cluster update, 
two Metropolis and eight over-relaxation sweeps were performed\cite{Landau}. 
The $K$ dependence of the fourth-order cumulant $U_{L}(K)$ was determined 
using the single-histogram re-weighting method\cite{Ferr}.
This method enables us to obtain accurate thermodynamic information over 
the entire scaling region using Monte Carlo simulations performed at only a few different values of
$K$.  We have performed Monte Carlo simulation on simple 
cubic lattices of size  $L\times L\times L$ with $6\leq L\leq 42$ using periodic boundary 
conditions applied in all directions and $10^{6}$ MC steps. 
We carried out of the order of 10000 thermalization steps and of
the order of 20000 measurements.
After we estimated the critical coupling $K_{c}$, we computed the magnetization
and the magnetic susceptibility at the critical coupling $K_{c}$.

\section{Results and Discussion}
\label{sec2}
In this section, we first have to determine the critical coupling $K_{c}$, and
then to examine the static behavior around $K_{c}$. Binder's fourth-order 
cumulant\cite{Binder} $U_{L}(K)$ is a convenient quantity that we use in order to estimate the critical 
coupling $K_{c}$ and the correlation length exponent $\nu$.

\begin{figure}[htp]
\epsfxsize=\figwidth\centerline{\epsffile{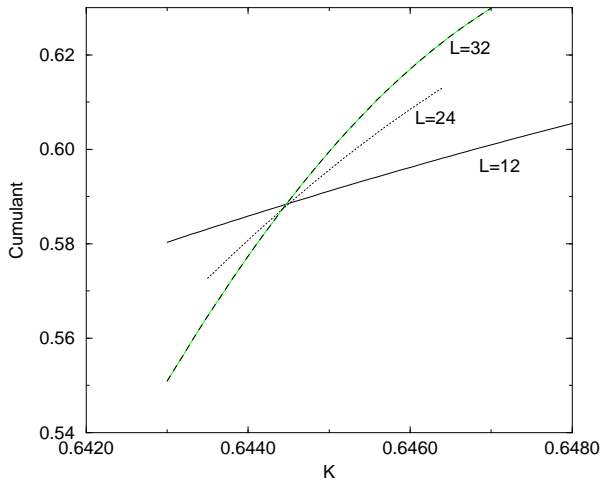}}
\caption{
Fourth-order cumulant $U_{L}(K)$ versus coupling $K$ for
lattice sizes $L$=12, 24, and 32.
}
\label{fi-1}
\end{figure}

Near the critical coupling $K_{c}$, the cumulant is expanded as
\begin{equation}
U_{L}=U^{\ast}+U_{1}L^{1/\nu}(1-\frac{T}{T_{c}})+\cdot\cdot\cdot\cdot.
\end{equation}
Therefore, if we plot $U_{L}(K)$ versus the coupling $K$ for several 
different sizes $L$, it is expected 
that the curves for different values of $L$ cross at the critical coupling $K_{c}$.
In order to find the $K$ dependence of the fourth-order cumulant $U_{L}(K)$,
we performed simulations for each lattice size from $L=6$ to $L=42$ 
at $K$=0.6450 which is chosen to be close to previous estimates for  the critical inverse temperature\cite{Ferer,Oh}.
The $U_{L}(K)$ curves were calculated from the histograms 
and are shown in Fig. \ref{fi-1} for $L$=12, 24, and 32.

\begin{figure}[htp]
\epsfxsize=\figwidth\centerline{\epsffile{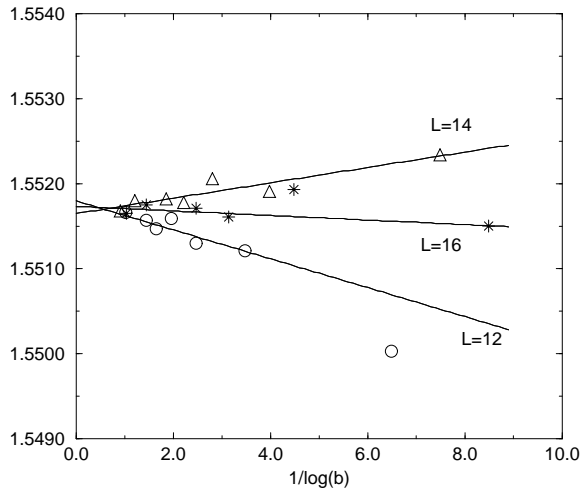}}
\caption{
Estimates for $T_{c}$ plotted versus inverse logarithm of the scale factor 
b=$L'/L$. The extrapolation leads to an estimate of $T_{c}$=1.5518(2).
}
\label{fi-2}
\end{figure}

If one wishes to obtain higher accurary, then one needs to examine Fig. 1 more 
carefully and to see that the  points where each pair of curves cross are slightly different 
for different pairs of lattices; in fact the points where the curves cross move slowly to 
lower couplings for larger system sizes. 
For the pair which corresponds to our largest lattice sizes $L$=32 and 42, the  point where they cross is
$K_{c}\approx 0.64455$.
In order to extract  more precise critical coupling $K_{c}$ from our data,
we compare the curves of $U_{L}$ for the two different lattice sizes $L$ and $L'=bL$ and then
find the location of the intersection of two different curves $U_{L}$ and $U_{L'}$.
As a result of the residual corrections to the finite size scaling \cite{Binder}, the locations
depend on the scale factor $b=L'/L$. We used the crossing points of the $L$=12, 14, and, 16
curves with all the other ones with higher $L'$ value respectively.
Hence we need to extrapolate the results of this method 
for (ln$b$)$^{-1} \longrightarrow 0$ using $(U_{bL}/U_{L})_{T=T_{c}}=1$.
In Fig. \ref{fi-2} we show the estimate for the critical temperature $T_{c}$. Our final estimate
for $T_{c}$ is  
\begin{equation}
T_{c}=1.5518(2),  K_{c}=0.6444(1).
\end{equation}
For comparison, the previous estimates are $T_{c}$=1.54(1)\cite{Costa,Oh} 
obtained using Monte Carlo simulation and $T_{c}$=1.552(3)\cite{Ferer}
obtained using high-temperature series. The latter result obtained with an 
expansion is surprisingly close to ours.

\begin{figure}[htp]
\epsfxsize=\figwidth\centerline{\epsffile{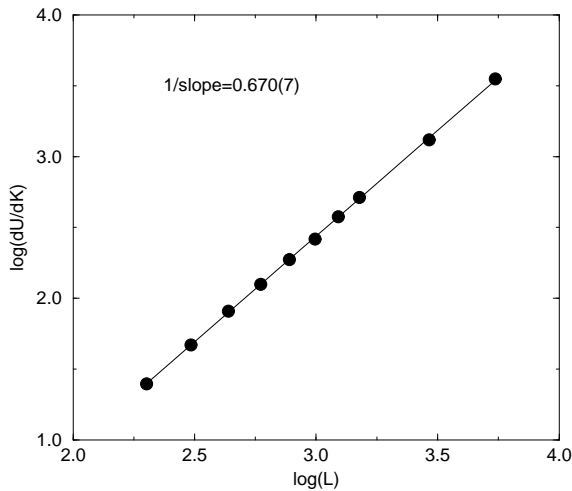}}
\caption{
Log-log plot of the slopes of $U$ near the crossing point versus $L$.
The slope gives an estimate for the critical exponent $\nu$=0.670(7).
}
\label{fi-3}
\end{figure}

In order to extract the critical exponent $\nu$, we performed finite-size scaling 
analysis of the slopes of $U_{L}$ versus $L$ near our estimated critical point $K_{c}$.
In the finite-size scaling region, the slope of the cumulant at $K_{c}$ varies
with system size like $L^{1/\nu}$,
\begin{equation}
\frac{dU_{L}}{dK} \sim L^{1/\nu}.
\end{equation} 
In Fig. \ref{fi-3}  we show results of a finite-size scaling analysis for the slope of the cumulant.
We obtained the value of the static exponent $\nu$,
\begin{equation}
\nu = 0.670(7).
\end{equation} 
For comparison, the field theoretical estimate\cite{Guillou} is $\nu$=0.669(2) and 
a recent experimental measurement gives $\nu$=0.6705(6)\cite{Goldner}.

\begin{figure}[htp]
\epsfxsize=\figwidth\centerline{\epsffile{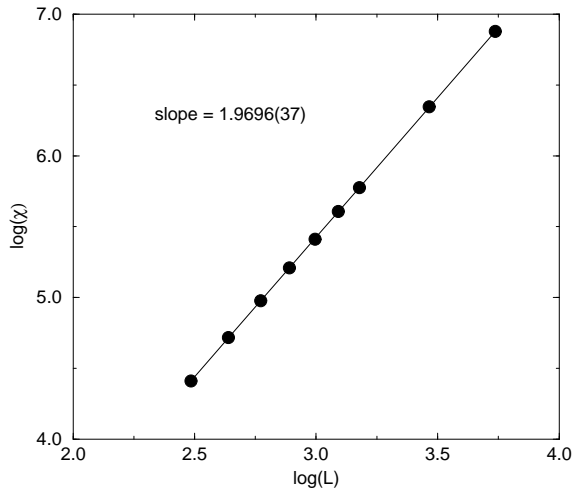}}
\caption{
Log-log plot of the susceptibility versus the lattice size $L$ at the critical coupling
$K_{c}$=0.6444. The slope gives an estimate for the critical exponent $\gamma/\nu$=1.9696(37).
}
\label{fi-4}
\end{figure}

In order to obtain the value of the exponent ratio $\gamma/\nu$, we calculated the magnetic 
susceptibility per spin $\chi$ at the critical coupling $K_{c}$.
The finite-size behavior for $\chi$ at the critical point is
\begin{equation}
\chi \sim L^{\gamma/\nu}.
\end{equation} 
Fig. \ref{fi-4} displays the finite-size scaling of the susceptibility 
$\chi$ calculated at $K_{c}$=0.6444. 
>From the log-log plot we obtained the value of the exponent 
ratio $\gamma/\nu$,
\begin{equation}
\gamma/\nu=1.9696(37). 
\end{equation}
>From the hyperscaling relation, $d\nu=\gamma+2\beta$, we get 
the exponent ratio $\beta/\nu$,
\begin{equation}
\beta/\nu=0.515(2).
\label{betanu}
\end{equation}

\begin{figure}[htp]
\epsfxsize=\figwidth\centerline{\epsffile{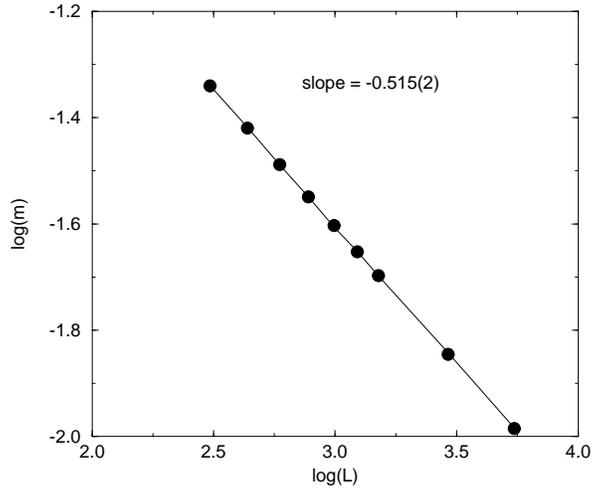}}
\caption{
Log-log plot of the magnetization versus the lattice size $L$ at the critical coupling
$K_{c}$=0.6444. The slope gives an estimate for the critical exponent $\beta/\nu$=0.515(3).
}
\label{fi-5}
\end{figure}

The equilibrium magnetization $m$ at $K_{c}$ 
should obey the relation
\begin{equation}
m \sim L^{-\beta/\nu}
\end{equation} 
for sufficiently larger $L$.
In Fig. \ref{fi-5} we show the results of a finite-size scaling analysis for the magnetization $m$.
We obtain the value of the exponent ratio $\beta/\nu$, 
\begin{equation}
\beta/\nu=0.515(2).
\end{equation}
This result agrees very closely to that of Eq. (\ref{betanu}) obtained from the susceptibility
and the fourth-order cumulant.

\begin{table}[htp] \centering
 \begin{tabular}{|l|l|l|} 
 \hline
 $L$ & \multicolumn{1}{c|}{$\chi$} & \multicolumn{1}{c|}{$m$} \\ \hline\hline
 12 &  82.39(28)   &  0.26195(55)  \\
 14 &  111.88(36)  &  0.24219(43)  \\
 16 &  145.12(59)  &  0.22567(55)  \\
 18 &  182.91(52)  &  0.21241(35)  \\
 20 &  224.08(85)  &  0.20072(49)  \\
 22 &  272.23(60)  &  0.19163(23)  \\
 24 &  322.35(98)  &  0.18308(32)  \\
 32 &  571.0(4.0)  &  0.15833(66)  \\
 42 &  972.0(4.8)  &  0.13749(40)  \\
 \hline
 \end{tabular}
\caption{\label{t1} Results for the magnetization and the susceptibility}
\end{table}

In conclusion, we determined the critical temperature and the exponents of the planar magnet model with 
three-component spins using a high-precision MC method, the single-histogram method, and  
the finite-size scaling theory. Our simulation results for the critical coupling and for the critical exponents
are  $K_{c}$=0.6444(1), $\nu$=0.670(7), $\gamma/\nu$=0.9696(37), and $\beta/\nu$=0.515(2).
Our calculated values for the critical temperature
and critical exponents are significantly more accurate that those previously calculated.
Comparison of our results with results of MC studies of the 3$D$ $XY$ model with two-component spins\cite{MC,S3,Janke,Hasenbusch}
shows that both the system with $\vec{S}_{i}=
(S^{x}_{i},S^{y}_{i})$ and the planar  magnet system with $\vec{S}_{i}=(S^{x}_{i},S^{y}_{i},S^{z}_{i})$
belong to the same universality class. 

\section{acknowledgements}

This work was supported by the National Aeronautics and Space
Administration under grant no. NAG3-1841.

\end{document}